\documentclass[12pt,prd,showpacs]{revtex4}
\usepackage{bm}
\usepackage{graphics}
\usepackage{rotating}
\usepackage{epsfig}
\begin{document}
\title{\begin{flushright}{\rm\normalsize HU-EP-02/01}\end{flushright}

Mass spectra of doubly heavy baryons in the relativistic quark model} 
\author{D. Ebert}
\affiliation{Institut f\"ur Physik, Humboldt--Universit\"at zu Berlin,
Invalidenstr.110, D-10115 Berlin, Germany}
\author{R. N. Faustov}
\author{V. O. Galkin}
\affiliation{Institut f\"ur Physik, Humboldt--Universit\"at zu Berlin,
Invalidenstr.110, D-10115 Berlin, Germany}
\affiliation{Russian Academy of Sciences, Scientific Council for
Cybernetics, Vavilov Street 40, Moscow 117333, Russia}
\author{A. P. Martynenko}
\affiliation{Samara State University, Pavlov Street 1, Samara 443011,
  Russia} 
\begin{abstract}
  Mass spectra of baryons consisting of two heavy ($b$ or $c$) and one
  light quarks are calculated in the framework of the relativistic
  quark model. The light quark-heavy diquark structure of the baryon
  is assumed. Under this assumption the ground and excited states of
  both the diquark and quark-diquark bound system are considered. The
  quark-diquark potential is constructed. The light quark is treated
  completely relativistically, while the expansion in the inverse heavy
  quark mass is used revealing the close similarity with the mass
  spectra of $B$ and $D$ mesons. We find that the relativistic
  treatment of the light quark plays an important role. The level
  inversion of the $p$-wave excitations of the light quark in doubly heavy
  baryons is discussed.  
\end{abstract}
\maketitle

\section{Introduction}

The description of baryons within the constituent quark model is a
very important problem in quantum chromodynamics (QCD). Since the
baryon is a three-body system
\cite{r}, its theory is much more complicated compared to the two-body
meson system. Up till now it is not even clear which of two main QCD
models, $Y$ law or $\Delta$ law, correctly describe the nonperturbative
(long-range) part of the quark interaction in the baryon
\cite{bali,ks}. The popular quark-diquark picture of a baryon is not
universal and does not work in all cases \cite{apkfl}. The success of
the heavy quark effective theory (HQET) \cite{mw} in predicting some
properties of the heavy-light $q\bar Q$ mesons ($B$ and $D$)
suggests to apply these methods to heavy-light baryons, too. The simplest
baryonic systems of this kind are the so-called doubly heavy baryons
($qQQ$) \cite{kkp,sw,bdgr,r,rdlp,efgms,gklo,kl}. The two heavy quarks
($b$ or $c$) compose in this case a bound diquark system in the
antitriplet colour state which serves as a localized colour
source. The light quark is orbiting around this heavy source at a
distance much larger ($\sim1/m_q$) than the source size
($\sim2/m_Q$). Thus the doubly heavy baryons look effectively like a
two-body bound system and strongly resemble the heavy-light $B$ and
$D$ mesons \cite{bali,isgur}. Then the HQET expansion in the inverse heavy
quark mass can be
used. The main distinction of the $qQQ$ baryon from the $q\bar Q$
meson is that the $QQ$ colour source though being almost localized
still is a composite system bearing integer spin values
($0,1,\dots$). Hence it follows that the interaction of the heavy
diquark with the light quark is not point-like but is smeared by the
form factor expressed through the overlap of the diquark wave
functions. Besides this the diquark excitations contribute to the
baryon excited states. 

In previous approaches for the calculation of doubly heavy baryon
masses the expansion in inverse powers not only of
the heavy quark mass ($m_Q$) but also  of the 
light quark mass  ($m_q$) is carried out. 
The estimates of the light quark velocity in these baryons show that
the light quark is highly relativistic ($v/c\sim 0.7\div 0.8$). Thus
the nonrelativistic approximation is not adequate for the light quark.
Here we present a consistent treatment of mass
spectra of the doubly heavy baryons in the framework of the
relativistic quark model based on the quasipotential wave equation
without employing the expansion in  $1/m_q$. Thus the light
quark is treated fully relativistically. Concerning the heavy 
diquark (quark) we apply the expansion in 
$1/M^d_{QQ}$ ($1/m_Q$).
We used a similar approach for
the calculation of the mass spectra of $B$ and $D$ mesons \cite{egf}. 

The paper is organized as follows. In Sec.~\ref{rqm} we describe our
relativistic quark model giving special emphasis to the construction
of the quark-quark interaction potential  in the diquark and the
quark-diquark interaction potential in the baryon. In Sec.~\ref{sec:dq} we
apply our model to the investigation of the heavy diquark
properties. The $cc$ and $bb$ diquark mass spectra are calculated. We
also determine the diquark interaction vertex with the gluon using the
quasipotential approach and calculated diquark wave functions. Thus we
take into account the internal structure of the diquark which considerably
modifies the quark-diquark potential at small distances and removes
fictitious singularities. In Sec.~\ref{sec:qp} we construct the
quasipotential of the interaction of a light quark with a heavy
diquark. The light quark is treated fully relativistically. We use the
expansion in inverse powers of the heavy diquark mass to simplify the
construction. First we consider the infinitely heavy diquark limit and
then include the $1/M_{QQ}^d$ corrections. In Sec.~\ref{sec:rd} we
present our predictions for the mass spectra of the ground and excited
states of $\Xi_{cc}$, $\Xi_{bb}$, $\Omega_{cc}$ and $\Omega_{bb}$
baryons. We consider both the excitations of the light quark and the heavy
diquark. The mixing between excited baryon states with the same total
angular momentum and parity is discussed. For  $\Xi_{cb}$ and
$\Omega_{cb}$ baryons, composed from heavy quarks of different
flavours we give predictions only for ground states, since the excited
states of the $cb$ diquark are unstable under the emission of soft gluons
\cite{gklo}.  Moreover, a detailed comparison of our predictions with other
approaches is given. We reveal the close similarity of the excitations
of the light quark in a doubly heavy baryon and a heavy-light
meson. We also test the fulfillment of different relations between
mass splittings of doubly heavy baryons with two $c$ or $b$
quarks as well as the relations between splittings in the doubly heavy
baryons and heavy-light mesons following from the heavy quark
symmetry. Section~\ref{sec:concl} contains our conclusions.

\section{Relativistic quark model}  
\label{rqm}

In the quasipotential approach and quark-diquark picture of doubly
heavy baryons the interaction of two heavy quarks in a diquark and the light
quark interaction with a heavy diquark in a baryon are described by the
diquark wave function ($\Psi_{d}$) of the bound quark-quark state
and by the baryon wave function ($\Psi_{B}$) of the bound quark-diquark
state respectively,  which satisfy the
quasipotential equation \cite{3} of the Schr\"odinger type \cite{4}
\begin{equation}
\label{quas}
{\left(\frac{b^2(M)}{2\mu_{R}}-\frac{{\bf
p}^2}{2\mu_{R}}\right)\Psi_{d,B}({\bf p})} =\int\frac{d^3 q}{(2\pi)^3}
 V({\bf p,q};M)\Psi_{d,B}({\bf q}),
\end{equation}
where the relativistic reduced mass is
\begin{equation}
\mu_{R}=\frac{E_1E_2}{E_1+E_2}=\frac{M^4-(m^2_1-m^2_2)^2}{4M^3},
\end{equation}
and $E_1$, $E_2$ are given by
\begin{equation}
\label{ee}
E_1=\frac{M^2-m_2^2+m_1^2}{2M}, \quad E_2=\frac{M^2-m_1^2+m_2^2}{2M},
\end{equation}
here $M=E_1+E_2$ is the bound state mass (diquark or baryon),
$m_{1,2}$ are the masses of heavy quarks ($Q_1$ and $Q_2$) which form
the diquark or of the heavy diquark ($d$) and light quark ($q$) which form
the doubly heavy baryon ($B$), and ${\bf p}$  is their relative momentum.  
In the center of mass system the relative momentum squared on mass shell 
reads
\begin{equation}
{b^2(M) }
=\frac{[M^2-(m_1+m_2)^2][M^2-(m_1-m_2)^2]}{4M^2}.
\end{equation}

The kernel 
$V({\bf p,q};M)$ in Eq.~(\ref{quas}) is the quasipotential operator of
the quark-quark or quark-diquark interaction. It is constructed with
the help of the
off-mass-shell scattering amplitude, projected onto the positive
energy states. In the following analysis we closely follow the
similar construction of the quark-antiquark interaction in mesons
which were extensively studied in our relativistic quark model
\cite{egf,mass}. For
the quark-quark interaction in a diquark we use the relation
$V_{QQ}=V_{Q\bar Q}/2$ arising under the assumption about the octet
structure of the interaction  from the difference of the $QQ$ and
$Q\bar Q$  colour states. An important role in this construction is
played by the Lorentz-structure of the confining  interaction. 
In our analysis of mesons while  
constructing the quasipotential of quark-antiquark interaction 
we adopted that the effective
interaction is the sum of the usual one-gluon exchange term with the mixture
of long-range vector and scalar linear confining potentials, where
the vector confining potential contains the Pauli terms.  
We use the same conventions for the construction of the quark-quark
and quark-diquark interactions in the baryon. The
quasipotential  is then defined by \cite{efgms,mass} 

(a) for the quark-quark ($QQ$) interaction
 \begin{equation}
\label{qpot}
V({\bf p,q};M)=\bar{u}_{1}(p)\bar{u}_{2}(-p){\cal V}({\bf p}, {\bf
q};M)u_{1}(q)u_{2}(-q),
\end{equation}
with
\[
{\cal V}({\bf p,q};M)=\frac23\alpha_SD_{ \mu\nu}({\bf
k})\gamma_1^{\mu}\gamma_2^{\nu}+\frac12 V^V_{\rm conf}({\bf k})
\Gamma_1^{\mu}({\bf k})\Gamma_{2;\mu}(-{\bf k})+
\frac12 V^S_{\rm conf}({\bf k}),
\]

(b) for quark-diquark ($qd$) interaction
\begin{eqnarray}
\label{dpot}
V({\bf p,q};M)&=&\frac{\langle d(P)|J_{\mu}|d(Q)\rangle}
{2\sqrt{E_d(p)E_d(q)}} \bar{u}_{q}(p)  
\frac43\alpha_SD_{ \mu\nu}({\bf 
k})\gamma^{\nu}u_{q}(q)\cr
&&+\psi^*_d(P)\bar u_q(p)J_{d;\mu}\Gamma_q^\mu({\bf k})
V_{\rm conf}^V({\bf k})u_{q}(q)\psi_d(Q)\cr 
&&+\psi^*_d(P)
\bar{u}_{q}(p)V^S_{\rm conf}({\bf k})u_{q}(q)\psi_d(Q), 
\end{eqnarray}
where $\alpha_S$ is the QCD coupling constant, the colour factor 
is equal to $2/3$ for quark-quark and $4/3$ for quark-diquark
interactions, $\langle d(P)|J_{\mu}|d(Q)\rangle$ is the vertex the
diquark-gluon interaction which is discussed in detail
below $\Big[$$P=(E_d,-{\bf p})$ and $Q=(E_d,-{\bf q})$,
$E_d=(M^2-m_q^2+M_d^2)/(2M)$ $\Big]$. $D_{\mu\nu}$ is the 
gluon propagator in the Coulomb gauge
\begin{equation}
D^{00}({\bf k})=-\frac{4\pi}{{\bf k}^2}, \quad D^{ij}({\bf k})=
-\frac{4\pi}{k^2}\left(\delta^{ij}-\frac{k^ik^j}{{\bf k}^2}\right),
\quad D^{0i}=D^{i0}=0,
\end{equation}
and ${\bf k=p-q}$; $\gamma_{\mu}$ and $u(p)$ are 
the Dirac matrices and spinors
\begin{equation}
\label{spinor}
u^\lambda({p})=\sqrt{\frac{\epsilon(p)+m}{2\epsilon(p)}}
\left(\begin{array}{c}
1\\ \displaystyle\frac{\mathstrut\bm{\sigma}\cdot{\bf p}}
{\mathstrut\epsilon(p)+m}
\end{array}\right)
\chi^\lambda,
\end{equation}
with $\epsilon(p)=\sqrt{{\bf p}^2+m^2}$. 

The diquark state in the confining part of the quark-diquark
quasipotential (\ref{dpot}) is described by the wave functions
\begin{equation}
  \label{eq:ps}
  \psi_d(p)=\left\{\begin{array}{ll}1 &\qquad \text{ for scalar diquark}\\
\varepsilon_d(p) &\qquad \text{ for (axial) vector diquark}
\end{array}\right. ,
\end{equation}
where the four vector
\begin{equation}\label{pv}
\varepsilon_d(p)=\left(\frac{(\bm{\varepsilon}_d\cdot {\bf
p})}{M_d},\bm{\varepsilon}_d+ \frac{(\bm{\varepsilon}_d\cdot {\bf p}){\bf
  p}}{M_d(E_d(p)+M_d)}\right), \qquad \varepsilon_d(p)\cdot p=0,  
\end{equation} 
is the polarization vector of the (axial) vector
diquark with momentum ${\bf p}$, $E_d(p)=\sqrt{{\bf p}^2+M_d^2}$ and
$\varepsilon_d(0)=(0,\bm{\varepsilon}_d)$ is the polarization vector in
the diquark rest frame. The effective long-range vector vertex of the
diquark can be presented in the form  
\begin{equation}
  \label{eq:jc}
  J_{d;\mu}=\left\{\begin{array}{ll}
  \frac{\displaystyle (P+Q)_\mu}{\displaystyle
  2\sqrt{E_d(p)E_d(q)}}&\qquad \text{ for scalar diquark}\cr
\frac{\displaystyle (P+Q)_\mu}{\displaystyle2\sqrt{E_d(p)E_d(q)}}
  -\frac{\displaystyle i\mu_d}{\displaystyle 2M_d}\Sigma_\mu^\nu 
\tilde k_\nu
  &\qquad \text{ for (axial) 
  vector diquark}\end{array}\right. ,
\end{equation}
where $\tilde k=(0,{\bf k})$ and we neglected the contribution of the
chromoquadrupole moment of
the (axial) vector diquark, which is suppressed by an additional power of
$k/M_d$. Here the antisymmetric tensor
\begin{equation}
  \label{eq:Sig}
  \left(\Sigma_{\rho\sigma}\right)_\mu^\nu=-i(g_{\mu\rho}\delta^\nu_\sigma
  -g_{\mu\sigma}\delta^\nu_\rho)
\end{equation}
and the (axial) vector diquark spin ${\bf S}_d$ is given by
$(S_{d;k})_{il}=-i\varepsilon_{kil}$. We choose the total
chromomagnetic moment of the (axial) vector 
diquark $\mu_d=2$ \cite{g2,efgms}.

The effective long-range vector vertex of the quark is
defined by \cite{egf,mass}
\begin{equation}
\Gamma_{\mu}({\bf k})=\gamma_{\mu}+
\frac{i\kappa}{2m}\sigma_{\mu\nu}\tilde k^{\nu}, \qquad \tilde
k=(0,{\bf k}),
\end{equation}
where $\kappa$ is the Pauli interaction constant characterizing the
anomalous chromomagnetic moment of quarks. In the configuration space
the vector and scalar confining potentials in the nonrelativistic
limit reduce to
\begin{eqnarray}
V^V_{\rm conf}(r)&=&(1-\varepsilon)V_{\rm conf}(r),\nonumber\\
V^S_{\rm conf}(r)& =&\varepsilon V_{\rm conf}(r),
\end{eqnarray}
with 
\begin{equation}
V_{\rm conf}(r)=V^S_{\rm conf}(r)+
V^V_{\rm conf}(r)=Ar+B,
\end{equation}
where $\varepsilon$ is the mixing coefficient. 

 All the parameters of
our model like quark masses, parameters of linear confining potential
$A$ and $B$, mixing coefficient $\varepsilon$ and anomalous
chromomagnetic quark moment $\kappa$ were fixed from the analysis of
heavy quarkonium masses \cite{mass} and radiative decays \cite{gf}. 
The quark masses
$m_b=4.88$ GeV, $m_c=1.55$ GeV, $m_s=0.50$ GeV, $m_{u,d}=0.33$ GeV and
the parameters of the linear potential $A=0.18$ GeV$^2$ and $B=-0.30$ GeV
have standard values of quark models.  The value of the mixing
coefficient of vector and scalar confining potentials $\varepsilon=-1$
has been determined from the consideration of the heavy quark expansion
\cite{fg} and meson radiative decays \cite{gf}.
Finally, the universal Pauli interaction constant $\kappa=-1$ has been
fixed from the analysis of the fine splitting of heavy quarkonia ${
}^3P_J$- states \cite{mass} and also from the heavy quark expansion
\cite{fg}. Note that the 
long-range  magnetic contribution to the potential in our model
is proportional to $(1+\kappa)$ and thus vanishes for the 
chosen value of $\kappa=-1$.

\section{Diquarks in the relativistic quark model}
\label{sec:dq}

The quark-quark interaction in the diquark consists of the sum of the
spin-independent and spin-dependent parts
\begin{equation}
  \label{eq:vqq}
 V_{QQ}(r)=V_{QQ}^{\rm SI}(r)+V_{QQ}^{\rm SD}(r). 
\end{equation}
The spin-independent part with the account of $v^2/c^2$ corrections
including retardation effects \cite{efghq} is given by
\begin{eqnarray}
\label{sipot}
V_{QQ}^{\rm SI}(r)&=&-\frac23\frac{\alpha_s(\mu^2)}{r} +\frac12(Ar+B) 
+\frac18\left(\frac{1}{m_1^2}+\frac{1}{m_2^2}\right) \Delta\left[
  -\frac23\frac{\alpha_s(\mu^2)}{r} +\frac12(1-\varepsilon)(1+2\kappa)
  Ar\right]\cr
&&+\frac{1}{2m_1m_2}\left\{-\frac23\frac{\alpha_s}{r}
\left[{\bf p}^2+\frac{({\bf p\cdot r})^2}{r^2}\right]\right\}_W
+\frac12\left[\frac{1-\varepsilon}{2m_1m_2}-\frac{\varepsilon}{4}
\left(\frac{1}{m_1^2}+\frac{1}{m_2^2}\right)\right]\cr
&&\times\left\{Ar\left[{\bf p}^2
-\frac{({\bf p\cdot r})^2}{r^2}\right]\right\}_W
+\frac12\left[\frac{1-\varepsilon}{2m_1m_2}-
 \frac{\varepsilon}{4}\left(\frac{1}{m_1^2}+
\frac{1}{m_2^2}\right)\right]B{\bf p}^2,
\end{eqnarray}
where $\{\dots\}_W$ denotes the Weyl ordering of operators and 
\begin{equation}
\label{alpha}
\alpha_s(\mu^2)=\frac{4\pi}{\beta_0\ln(\mu^2/\Lambda^2)},
\qquad \beta_0=11-\frac23 n_f,
\end{equation}
with $\mu$ fixed to be equal to the reduced mass, $n_f$ is a number of
flavours and $\Lambda=85$~MeV.

The spin-dependent part of the quark-quark potential can be presented
in our  model \cite{mass} as follows:
\begin{equation}
\label{vsd}
V_{QQ}^{\rm SD}(r)= a\ {\bf L}\cdot{\bf\tilde S}+b\left[\frac{3}{r^2}({\bf
    S}_1\cdot {\bf r})
({\bf S}_2\cdot {\bf r})-({\bf S}_1\cdot {\bf S}_2)\right] +c\ {\bf
    S}_1\cdot {\bf S}_2 +d\ {\bf L}\cdot({\bf S}_1-{\bf S}_2) , 
\end{equation}
\begin{eqnarray}
\label{a}
a&=& \frac{1}{m_1m_2}\Biggl\{\left(1+\frac{m_1^2+m_2^2}
  {4m_1m_2}\right)\frac23
  \frac{\alpha_s(\mu^2)}{r^3}-\frac12\frac{m_1^2+m_2^2}{4m_1m_2}\frac{A}{r}+
  \frac12(1+\kappa)\frac{(m_1+m_2)^2}{2m_1m_2}(1-\varepsilon)\frac{A}{r}
  \Biggr\},\cr
\label{b}
b&=& \frac{1}{3m_1m_2}\Biggl\{\frac{2\alpha_s(\mu^2)}{r^3}
+\frac12(1+\kappa)^2(1-\varepsilon)\frac{A}{r}\Biggr\},\cr
\label{c}
c&=& \frac{2}{3m_1m_2}\Biggl\{\frac{8\pi\alpha_s(\mu^2)}{3}\delta^3(r)
+\frac12(1+\kappa)^2(1-\varepsilon)\frac{A}{r}\Biggr\},\cr
d&=&\frac1{m_1m_2}\Biggl\{\frac{m_2^2-m_1^2}{4m_1m_2}\left[\frac23
\frac{\alpha_s(\mu^2)}{r^3}-\frac12\frac{A}{r}\right]+\frac12(1+\kappa)
\frac{m_2^2-m_1^2}{2m_1m_2}(1-\varepsilon)\frac{A}{r}
\Biggr\}, 
\end{eqnarray}
where ${\bf L}$ is the orbital momentum and ${\bf S}_{1,2}$, ${\bf
\tilde  S}={\bf S}_1+ {\bf S}_2$ are the spin momenta.

Now we can calculate the mass spectra of heavy 
diquarks with the account
of all relativistic corrections (including retardation effects) of
order $v^2/c^2$. For this purpose we substitute the quasipotential
which is a sum of the spin-independent  (\ref{sipot}) and
spin-dependent (\ref{vsd}) parts into the quasipotential equation
(\ref{quas}). Then we multiply the resulting expression from the left
by the quasipotential wave function of a bound state and
integrate with respect to the relative momentum. Taking into account
the accuracy of the calculations, we can use for the resulting matrix
elements the wave functions of Eq.~(\ref{quas}) with the static potential 
\begin{equation}
V_{QQ}^{\rm NR}(r)=-\frac23\frac{\alpha_s(\mu^2)}{r} +\frac12(Ar+B).
\end{equation}
As a result we obtain the mass formula 
\begin{eqnarray}
\label{mform}
\frac{b^2(M)}{2\mu_R}&=&W+\langle a\rangle\langle{\bf
  L}\cdot{\bf\tilde S}\rangle 
+\langle b\rangle \langle\left[\frac{3}{r^2}
({\bf S}_1\cdot {\bf r})
({\bf S}_2\cdot {\bf r})-({\bf S}_1\cdot {\bf S}_2)\right] \rangle
+\langle c\rangle \langle{\bf S}_1\cdot {\bf S}_2\rangle\cr
&&+\langle d\rangle\ \langle{\bf L}\cdot({\bf S}_1-{\bf S}_2)\rangle,
\end{eqnarray}
where
\begin{eqnarray}
W&=&\langle V_{QQ}^{\rm SI}\rangle+\frac{\langle {\bf p}^2\rangle}{2\mu_R},\cr
\langle{\bf L}\cdot{\bf\tilde S}\rangle&=&
\frac12(\tilde J(\tilde J+1)-L(L+1)-\tilde S(\tilde S+1)),\cr
\langle\left[\frac{3}{r^2}
({\bf S}_1\cdot {\bf r})
({\bf S}_2\cdot {\bf r})-({\bf S}_1\cdot {\bf S}_2)\right] \rangle&=&
-\frac{6(\langle{\bf L}\cdot{\bf\tilde S}\rangle)^2+3\langle{\bf L}
  \cdot{\bf\tilde S}\rangle -2\tilde S(\tilde
  S+1)L(L+1)}{2(2L-1)(2L+3)},\cr 
\langle{\bf S}_1\cdot {\bf S}_2\rangle&=&\frac12\left(\tilde S(\tilde
  S+1)-\frac32\right),
\qquad
{\bf \tilde S}={\bf S}_1+{\bf S}_2,\nonumber
\end{eqnarray}
and $\langle a\rangle$, $\langle b\rangle$, $\langle c\rangle$,
$\langle d\rangle$ are the appropriate
averages over radial wave functions of Eq.~(\ref{a}).  We
use the notations for the heavy diquark classification:
$n^{2\tilde S+1}L_{\tilde J}$,  where $n=1,2,\dots$ is  a radial
quantum number, $L$ is the angular momentum, $\tilde S=0,1$ 
is the total spin of two heavy quarks, and $\tilde J=L-\tilde S, L,
L+\tilde S$  is  the
total angular momentum (${\bf\tilde J}={\bf L}+{\bf\tilde S}$), which is
considered as the spin of diquark (${\bf S}_d$) in the following section.
The first term on the right-hand side of the mass formula
(\ref{mform}) contains all spin-independent contributions, the second
and the last terms describe the spin-orbit interaction, the third term
is responsible for the tensor interaction, while the
forth term gives the spin-spin interaction. The results of our
numerical calculations of the mass spectra of $cc$ and
$bb$ diquarks are presented in Tables~\ref{tab:dcc} and \ref{tab:dbb}.
The mass of the ground state of the $bc$ diquark in axial vector ($1{}^3S_1$)
state is $$M_{bc}^{A}=6.526~{\rm GeV}$$ and in scalar ($1{}^1S_1$) state is
$$M_{bc}^{S}=6.519~{\rm GeV}.$$   

\begin{table}[htbp]
    \caption{Mass spectrum and mean squared radii of the $cc$ diquark.}
    \label{tab:dcc}
    \begin{ruledtabular}
    \begin{tabular}{cccccc}
State & Mass &$\langle r^2\rangle^{1/2}$ & State & Mass & $\langle
r^2\rangle^{1/2}$\\ 
&(GeV)& (fm)& &(GeV)& (fm)\\
\hline
$1{}^3S_1$& 3.226&0.56 & $1{}^1P_1$ & 3.460&0.82\\
$2{}^3S_1$& 3.535&1.02 & $2{}^1P_1$ & 3.712&1.22\\
$3{}^3S_1$& 3.782&1.37 & $3{}^1P_1$ & 3.928&1.54\\
    \end{tabular}
    \end{ruledtabular}
\end{table}

\begin{table}[htbp]
    \caption{Mass spectrum and mean squared radii of $bb$ diquark.}
    \label{tab:dbb}
    \begin{ruledtabular}
    \begin{tabular}{cccccc}
State & Mass&$\langle r^2\rangle^{1/2}$  & State & Mass &$\langle
r^2\rangle^{1/2}$\\ 
&(GeV)& (fm)& &(GeV)& (fm)\\
\hline
$1{}^3S_1$& 9.778&0.37 & $1{}^1P_1$ & 9.944&0.57\\
$2{}^3S_1$& 10.015&0.71 & $2{}^1P_1$ & 10.132&0.87\\
$3{}^3S_1$& 10.196&0.98 & $3{}^1P_1$ & 10.305&1.12\\
$4{}^3S_1$& 10.369&1.22 & $4{}^1P_1$ & 10.453&1.34\\
\end{tabular}
\end{ruledtabular}
\end{table}

In order to determine the diquark interaction with the gluon field, which
takes into account the diquark structure, it is
necessary to calculate the corresponding matrix element of the quark
current between diquark states. This diagonal matrix element can be
parametrized by the following set of elastic form factors

(a) scalar diquark ($S$)
\begin{equation}
  \label{eq:sff}
  \langle S(P)\vert J_\mu \vert S(Q)\rangle=h_+(k^2)(P+Q)_\mu,
\end{equation}

(b) (axial) vector diquark ($V$) 
\begin{eqnarray}
  \label{eq:avff}
\langle V(P)\vert J_\mu \vert V(Q)\rangle&=&
-[\varepsilon_d^*(P)\cdot\varepsilon_d(Q)]h_1(k^2)(P+Q)_\mu\cr
&&+h_2(k^2)
\left\{[\varepsilon_d^*(P) \cdot Q]\varepsilon_{d;\mu}(Q)+
  [\varepsilon_d(Q) \cdot P] 
\varepsilon^*_{d;\mu}(P)\right\}\cr
&&+h_3(k^2)\frac1{M_{V}^2}[\varepsilon^*_d(P) \cdot Q]
    [\varepsilon_d(Q) \cdot P](P+Q)_\mu, 
\end{eqnarray}
where $k=P-Q$ and $\varepsilon_d(P)$ is the polarization vector of the
(axial) vector diquark (\ref{pv}). 

\begin{figure}
\includegraphics{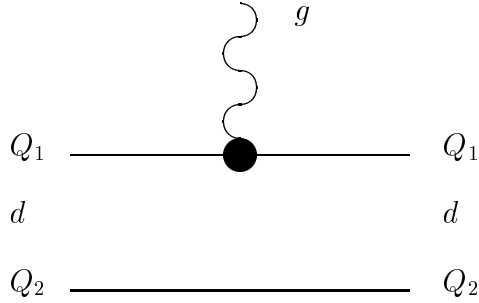}
  \caption{\label{fig:pic1} Lowest order vertex function $\Gamma^{(1)}$
corresponding to Eq.~(\ref{gam1}). The gluon interaction only with one
heavy quark is shown.}
\end{figure}

In the quasipotential approach,  the
matrix element of the quark current $J_\mu=\bar Q
\gamma^\mu Q$  between the diquark states ($d$) has the form \cite{f}
\begin{equation}\label{mxet}
\langle d(P) \vert J_\mu (0) \vert d(Q)\rangle
=\int \frac{d^3p\, d^3q}{(2\pi )^6} \bar \Psi^{d}_P({\bf
p})\Gamma _\mu ({\bf p},{\bf q})\Psi^{d}_Q({\bf q}),\end{equation}
where $\Gamma _\mu ({\bf p},{\bf
q})$ is the two-particle vertex function and  $\Psi^{d}_P$ are the
diquark wave functions projected onto the positive energy states of
quarks and boosted to the moving reference frame with momentum $P$.
The leading contribution to $\Gamma$ comes from Fig.~\ref{fig:pic1}.
Other nonleading terms are the consequence of the
projection onto the positive-energy states and give contributions
only at order of $1/m_Q^2$ for diquarks composed
from two heavy quarks and can be neglected. Thus we will limit our
analysis only to the
contribution coming from Fig.~\ref{fig:pic1}. The corresponding vertex
function is given by
\begin{equation}\label{gam1}
\Gamma_\mu ^{(1)}({\bf p},{\bf q})=\bar
u_{Q_1}(p_1)\gamma^\mu u_{Q_1}(q_1)(2\pi)^3\delta({\bf p}_2-{\bf q}_2)+
(1\leftrightarrow 2),
\end{equation} 
where \cite{f} 
\begin{eqnarray*} 
p_{1,2}&=&\epsilon_{1,2}(p)\frac{P}{M_{d}}
\pm\sum_{i=1}^3 n^{(i)}(P)p^i,\cr
q_{1,2}&=&\epsilon_{1,2}(q)\frac{Q}{M_{d}} \pm \sum_{i=1}^3 n^{(i)}
(Q)q^i,
\end{eqnarray*}
and $n^{(i)}$ are three four-vectors defined by
$$ n^{(i)\mu}(P)=\left\{ \frac{P^i}{M},\ \delta_{ij}+
\frac{P^iP^j}{M(E+M)}\right\}, \quad E=\sqrt{{\bf P}^2+M^2}.$$
 
\begin{figure}
\includegraphics{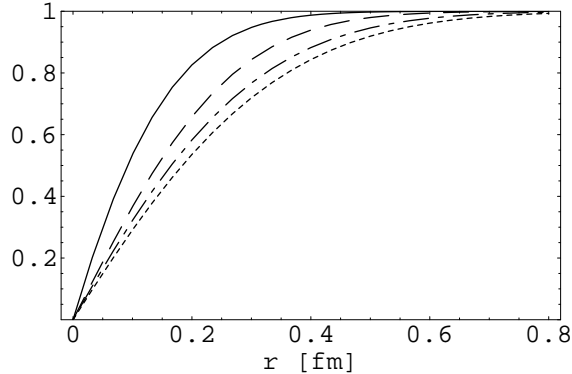}
\caption{\label{fig:ff} The form factors $F(r)$ for the 
$cc$ diquark. The solid curve is for the $1S$ state, the dashed curve
for the $1P$ state, the dashed-dotted curve for the $2S$ state, and the dotted curve
for the $2P$ state.}  
\end{figure}

We substitute the vertex function $\Gamma^{(1)}$ given by
Eq.~(\ref{gam1}) in the matrix element (\ref{mxet}) and expand it in
$1/m_Q$ up to the leading order. Comparing the resulting expressions with
the form factor decompositions (\ref{eq:sff}) and (\ref{eq:avff}) we
find 
\begin{eqnarray}
  \label{eq:hf}
  h_+(k^2)&=&h_1(k^2)=h_2(k^2)=2F({\bf k}^2),\cr
h_3(k^2)&=&0,\cr
F({\bf k}^2)&=&\frac{\sqrt{E_{d}M_{d}}}{E_{d}+M_{d}}
  \left[\int \frac{d^3p}{(2\pi )^3} \bar\Psi_{d}
\left({\bf p}+
\frac{2m_{Q_2}}{E_{d}+M_{d}}{\bf k } \right)\Psi_{d}({\bf
  p})+(1\leftrightarrow 2)\right],
\end{eqnarray}
where $\Psi_{d}\equiv \Psi^{d}_0$ are the diquark wave functions at rest.
We calculated corresponding form factors $F(r)/r$ which are the Fourier
transforms of $F({\bf k}^2)/{\bf k}^2$ using the diquark wave
functions found 
by numerical solving the quasipotential equation. In Fig.~\ref{fig:ff}
the functions $F(r)$ for the  $cc$ diquark in ($1S,1P,2S,2P$) states
are shown as 
an example. We see that the slope of $F(r)$ decreases with the
increase of the diquark excitation.
Our estimates show that this form factor can be approximated  with a
high accuracy by the expression 
\begin{equation}
  \label{eq:fr}
  F(r)=1-e^{-\xi r -\zeta r^2},
\end{equation}
which agrees with previously used approximations \cite{ms}.
The values of parameters $\xi$ and $\zeta$ for different $cc$ and $bb$
diquark states are given in Tables~\ref{tab:fcc} and \ref{tab:fbb}.
As we see the functions $F(r)$ vanish in the limit $r\to 0$ and become
unity for large values of $r$. Such a behaviour can be easily understood
intuitively. At large distances a diquark can be well approximated by a
point-like object and its internal structure cannot be resolved. When
the distance to the diquark decreases the internal structure plays a more
important role. As the distance approaches zero, the interaction weakens and
turns to zero for $r=0$ since this point coincides with the center of
gravity of the two heavy quarks forming the diquark. Thus the function $F(r)$
gives an important contribution to the short-range part of the
interaction of the light quark with the heavy diquark in the baryon and
can be neglected for the long-range (confining) interaction. It is
important to note that the inclusion of such a function removes a
fictitious singularity $1/r^3$ at the origin arising from the
one-gluon exchange part of the quark-diquark potential when the
expansion in inverse powers of the heavy-quark is used.        

\begin{table}
\caption{\label{tab:fcc}Parameters  $\xi$ and $\zeta$ for ground and
  excited states  of $cc$ diquark.}
\begin{ruledtabular}
\begin{tabular}{cccccc}
State & $\xi$  & $\zeta$  & State & $\xi$ & $\zeta$  \\
&(GeV)&(GeV$^2$)& &(GeV)&(GeV$^2$)\\
\hline
$1S$ & 1.30 & 0.42 & $1P$ & 0.74 & 0.315\\
$2S$ & 0.67 & 0.19 & $2P$ & 0.60 & 0.155\\
$3S$ & 0.57 & 0.12 & $3P$ & 0.55 & 0.075\\
\end{tabular}
\end{ruledtabular}
\end{table}
\begin{table}
\caption{\label{tab:fbb}Parameters  $\xi$ and $\zeta$ for ground and
  excited states  of $bb$ diquark.}
\begin{ruledtabular}
\begin{tabular}{cccccc}
State & $\xi$ & $\zeta$ & State & $\xi$ & $\zeta$ \\
&(GeV)&(GeV$^2$)& &(GeV)&(GeV$^2$)\\
\hline
$1S$ & 1.30 & 1.60 & $1P$ & 0.90 & 0.59\\
$2S$ & 0.85 & 0.31 & $2P$ & 0.65 & 0.215\\
$3S$ & 0.66 & 0.155 & $3P$ & 0.58 & 0.120\\
$4S$ & 0.56 & 0.09 & $4P$ & 0.51 & 0.085\\
\end{tabular}
\end{ruledtabular}
\end{table}

\section{Quasipotential of the interaction of a light quark  with a heavy
diquark} 
\label{sec:qp}
The expression for the quasipotential (\ref{dpot}) can, in principle, be
used for arbitrary quark and diquark masses.  The substitution
of the Dirac spinors (\ref{spinor}) and diquark form factors
(\ref{eq:sff}) and (\ref{eq:avff}) into (\ref{dpot}) results in an extremely
nonlocal potential in the configuration space. Clearly, it is very hard to 
deal with such potentials without any simplifying expansion.
Fortunately, in the case of heavy-diquark light-quark picture of the
baryon, one can carry out (following HQET)
the expansion in inverse powers of the heavy diquark mass
$M_{d}$. The leading terms then follow
in the limit $M_{d}\to \infty$. 

\subsection{Infinitely heavy diquark limit}

In the limit $M_{d}\to\infty$ the heavy diquark vertices (\ref{eq:sff})
and (\ref{eq:avff}) have only the zeroth component, and the diquark mass
and spin decouple from the consideration. As a result we get in this
limit the quasipotential for the light quark similar to the one in
the heavy-light meson in the
limit of an infinitely heavy antiquark \cite{egf}. The only difference
consists in the extra factor $F(k^2)$, defined in (\ref{eq:hf}), in the
one-gluon exchange part which accounts for the heavy diquark
structure. The quasipotential in this limit is given by  
\begin{eqnarray}
\label{ipot}
V({\bf p},{\bf q};M)&=& \bar u_q(p)\Bigg\{-\frac43\alpha_s F(k^2)
\frac{4\pi}{{\bf k}^2}\gamma_q^0\cr
& & +V_{\rm conf}^V({\bf k})\left[\gamma_q^0+\frac{\kappa}{2m_q}
\gamma_q^0(\bm{\gamma}{\bf k})\right]+V_{\rm conf}^S({\bf k})
\Bigg\}u_q(q).
\end{eqnarray}
The resulting interaction is still nonlocal in configuration
space. However, taking into account that doubly heavy baryons are
weakly bound, we can replace $\epsilon_q(p) \to E_q=(M^2-M_d^2+m_q^2)/(2M)$ 
in  the Dirac spinors (\ref{spinor}) \cite{egf}. 
Such simplifying substitution is widely used
in quantum electrodynamics \cite{bs,mf,km} and introduces only minor 
corrections of order of the ratio of the binding energy $\langle V
\rangle$ to $E_q$. This substitution makes the Fourier 
transformation of the potential (\ref{ipot}) local. In contrast with
the heavy-light meson case no special consideration of the one-gluon
exchange term is necessary, since the presence of the diquark structure
described by an extra function $F(k^2)$ in Eq.~(\ref{ipot}) removes
fictitious $1/r^3$ singularity at the origin in configuration space.

 The resulting local quark-diquark potential for $M_d\to \infty$
can be presented in configuration space in the following form
\begin{eqnarray}
\label{vinf}
\!\!\!\!\!\! V_{M_d\to \infty}(r)&=& \frac{E_q+m_q}{2E_q}\Bigg[V_{\rm Coul}(r)
+V_{\rm conf}(r) + \frac{1}{(E_q+m_q)^2}\Bigg\{{\bf p}[ 
V_{\rm Coul}(r)\cr
& & +V_{\rm conf}^V(r)-V_{\rm conf}^S(r)]{\bf p}
-\frac{E_q+m_q}{2m_q}\Delta V_{\rm
conf}^V(r)[1-(1+\kappa)] \cr
& & +\frac{2}{r}\left(V_{\rm Coul}'(r)-V_{\rm conf}'^S(r) -
V_{\rm conf}'^V(r)\left[\frac{E_q}{m_q}
-2(1+\kappa)\frac{E_q+m_q}{2m_q}\right]\right)
{\bf l}\cdot{\bf S}_q\Bigg\}\Bigg],
\end{eqnarray}
where $V_{\rm Coul}(r)=-(4/3)\alpha_s F(r)/r$ is the smeared Coulomb
potential. The prime denotes differentiation with respect to $r$,
${\bf l}$ is the orbital momentum, 
and ${\bf S}_q$ is the spin operator of the light quark. Note that
the last term in (\ref{vinf}) is of the same order as the first two
terms and thus cannot be treated perturbatively. It is important to
note that  the quark-diquark potential $V_{M_d\to \infty}(r)$
almost coincides with the quark-antiquark potential in heavy-light
($B$ and $D$)  mesons for $m_Q\to \infty$ \cite{egf}. The only
difference is the presence of the extra factor $F(r)$ in $V_{\rm Coul}(r)$
which accounts for the internal structure of the diquark. This is the
consequence of the heavy quark (diquark) limit in which its spin and
mass decouple from the consideration. 

In the infinitely heavy diquark limit the quasipotential equation
(\ref{quas}) in configuration space becomes
\begin{equation}
\label{iquas}
\left(\frac{E_q^2-m_q^2}{2E_q}-\frac{{\bf p}^2}{2E_q}\right)
\Psi_B(r)=V_{M_d\to\infty}(r)\Psi_B(r),
\end{equation}
and the mass of the baryon is given by $M=M_{d}+E_q$.

Solving (\ref{iquas}) numerically  we 
get the eigenvalues $E_q$ and the baryon wave functions $\Psi_B$. The
obtained  results are presented in Table~\ref{eq}.  We use the
notation $n_{d}L n_{q}l(j)$ for the classification of baryon states in
the infinitely heavy diquark limit. Here we first give  the radial
quantum number $n_{d}$  and the angular momentum $L$  of the heavy
diquark. Then the radial quantum number  $n_q$, the angular momentum
$l$ and  the value $j$ of the total angular momentum (${\bf j}={\bf
  l}+{\bf S}_q$) of the light quark are shown.
We see that the heavy diquark spin and mass decouple in  the limit
$M_{d}\to\infty$,  and thus we get the number of degenerated states
in accord with the heavy quark symmetry prediction. This symmetry
predicts also an almost equality of corresponding light-quark energies
$E_q$ for $bbq$ and $ccq$ baryons and their nearness in the same limit
to the light-quark energies $E_q$
of $B$ and $D$ mesons \cite{egf}. The small deviations of the baryon
energies from values of the meson energies are connected with the
different forms of the singularity smearing at $r=0$ in the baryon and
meson cases.   
\begin{table}
\caption{\label{eq}The values of $E_q$ in the limit $M_{d}\to\infty$ 
(in GeV).}
\begin{ruledtabular}
\begin{tabular}{ccccc}
Baryon &{$ccq$}&{$ccs$}
&{$bbq$}&{$bbs$}\\
State& $E_{q}$ & $E_s$ & $E_{q}$& $E_s$ \\
\hline
$1S1s(1/2)$& 0.491 & 0.638 & 0.492 & 0.641\\ 
$1S1p(3/2)$& 0.788 &0.906& 0.785 & 0.904 \\
$1S1p(1/2)$& 0.877 &0.968& 0.880 & 0.969 \\
$1S2s(1/2)$& 0.987 &1.080& 0.993& 1.084\\
$1P1s(1/2)$& 0.484 & 0.633 & 0.489 & 0.636\\ 
$1P1p(3/2)$& 0.793 &0.909& 0.789 & 0.906 \\
$1P1p(1/2)$& 0.873 &0.965& 0.876 & 0.967 \\
$1P2s(1/2)$& 0.980 &1.075& 0.984& 1.078\\
$2S1s(1/2)$& 0.481 & 0.631 & 0.486 & 0.634\\ 
$2S1p(3/2)$& 0.794 &0.909& 0.791 & 0.908 \\
$2S1p(1/2)$& 0.871 &0.963& 0.874 & 0.965 \\
$2S2s(1/2)$& 0.979 &1.074& 0.982& 1.076\\
$2P1s(1/2)$& 0.479 &0.630& 0.481& 0.631\\
$3S1s(1/2)$& 0.478 & 0.630 & 0.480 & 0.630\\
\end{tabular}
\end{ruledtabular}
\end{table}

\subsection{$1/M_{d}$ corrections}

The heavy quark symmetry degeneracy of states is broken by $1/M_{d}$
corrections. The corrections of order $1/M_{d}$ to the potential
(\ref{vinf}) arise from the spatial components of the heavy diquark
vertex.  Other contributions at first
order in $1/M_{d}$ come from the one-gluon-exchange potential and the
vector confining potential, while the scalar potential gives no 
contribution at first order. The resulting $1/M_{d}$ correction to  
the quark-diquark potential (\ref{vinf}) is given by the following
expression

(a) scalar diquark
\begin{eqnarray}
\label{svcor}
\delta V_{1/M_{d}}(r)&=&\frac{1}{E_qM_{d}}\Bigg\{{\bf p}\left[V_{\rm
Coul}(r)+V^V_{\rm conf}(r)\right]{\bf p}
+V'_{\rm Coul}(r)\frac{{\bf l}^2}{2r}\cr
&&-\frac{1}{4}\Delta V^V_{\rm conf}(r)+\left[\frac{1}{r}V'_{\rm 
Coul}(r)+\frac{(1+\kappa)}{r}V'^V_{\rm conf}(r)\right]{\bf l\cdot S}_q
\Bigg\},
\end{eqnarray}

(b) (axial) vector diquark
\begin{eqnarray}
\label{avcor}
\delta V_{1/M_{d}}(r)&=&\frac{1}{E_qM_{d}}\Bigg\{{\bf p}\left[V_{\rm
Coul}(r)+V^V_{\rm conf}(r)\right]{\bf p}
+V'_{\rm Coul}(r)\frac{{\bf l}^2}
{2r}\cr
&&-\frac{1}{4}\Delta V^V_{\rm conf}(r)+\left[\frac{1}{r}V'_{\rm 
Coul}(r)+\frac{(1+\kappa)}{r}V'^V_{\rm conf}(r)\right]{\bf l\cdot S}_q\cr
&&+\frac12\left[\frac{1}{r}V'_{\rm 
Coul}(r)+\frac{(1+\kappa)}{r}V'^V_{\rm conf}(r)\right]{\bf l\cdot S}_{d}\cr
& & +\frac13\bigglb(\frac{1}{r}V'_{\rm Coul}(r)-V''_{\rm Coul}(r)
+(1+\kappa)\left[\frac{1}{r}V'^V_{\rm conf}(r)-V''^V_{\rm conf}(r)
\right]\biggrb)\cr
&&\times\left[-{\bf S}_q\cdot{\bf S}_{d}+\frac{3}{r^2}({\bf S}_q\cdot{\bf
r})({\bf S}_{d}\cdot{\bf r})\right]\cr
& &+\frac23\left[\Delta V_{\rm Coul}(r)+(1+\kappa)\Delta V^V_{\rm 
conf}(r)\right]{\bf S}_{d}\cdot{\bf S}_q\Bigg\},
\end{eqnarray}
where ${\bf S}={\bf S}_q+{\bf S}_{d}$ is the total spin, ${\bf S}_{d}$
is the diquark spin (which is equal to the total angular momentum
${\bf \tilde J}$ of two heavy quarks forming the diquark). The first
three terms in (\ref{avcor}) represent
spin-independent  corrections, the fourth and the fifth 
terms are responsible for the spin-orbit interaction, the sixth one is 
the tensor interaction and the last one is the spin-spin interaction.
It is necessary to note that the confining vector interaction gives a
contribution to the spin-dependent part which is proportional to
$(1+\kappa)$. Thus it vanishes for the chosen value
of $\kappa=-1$, while the confining vector contribution to 
the spin-independent part is nonzero. 
  
In order to estimate the matrix elements of spin-dependent terms in the
$1/M_d$ corrections to the quark-diquark potential (\ref{svcor}) and
(\ref{avcor}) as well as different mixings of baryon states, it is
convenient to use the following relations  
\begin{equation}
  \label{eq:jjq}
  |J;j\rangle=\sum_S(-1)^{J+l+S_d+S_q}\sqrt{(2S+1)(2j+1)}\left\{
   {S_d \atop l}\ {S_q\atop J}\ {S\atop j}\right\} |J,S\rangle
\end{equation}
and
\begin{equation}
  \label{eq:jjjd}
  |J;j\rangle=\sum_{J_d}(-1)^{J+l+S_d+S_q}\sqrt{(2J_d+1)(2j+1)}\left\{
   {S_d \atop S_q}\ {l\atop J}\ {J_d\atop j}\right\} |J,J_d\rangle,
\end{equation}
where ${\bf J}={\bf j}+{\bf S}_d$ is the baryon total angular
momentum, ${\bf j}={\bf l}+{\bf S}_q$ is the
light quark total angular momentum, ${\bf S}={\bf S}_q+{\bf S}_d$ is
the baryon total spin, and ${\bf J}_d={\bf L}+{\bf S}_d$.

\section{Results and discussion}
\label{sec:rd}

For the description of the quantum numbers of baryons we use the
notations $(n_dLn_ql)J^P$, where we first show the radial quantum
number of the diquark ($n_d=1,2,3\dots$) and its orbital momentum by a
capital letter ($L=S,P,D\dots$) , then the radial quantum number of the
light quark ($n_q=1,2,3\dots$) and its orbital momentum by a
lowercase letter ($l=s,p,d\dots$), and at the end the total angular
momentum $J$ and parity $P$ of the baryon.  

The presence of the spin-orbit interaction proportional to ${\bf l
\cdot S}_{d}$ and of the tensor interaction  in the quark-diquark potential 
at $1/M_d$ order (\ref{avcor}) results in a mixing of states which have
the same total angular momentum $J$ and parity but different light
quark total momentum $j$. For example, the baryon states with diquark
in the ground state and light quark in the $p$-wave ($1S1p$) for $J=1/2$
or $3/2$ have different values of the light quark angular momentum $j=1/2$
and $3/2$, which mix between themselves. In the case of the $ccq$ baryon
we have the mixing matrix for $J=1/2$
\begin{equation}
  \label{eq:mmc12}
  \left(
    \begin{array}{cc}
-55.6& -7.3\\ -8.5&-37.9
    \end{array}\right)\ {\rm MeV},
\end{equation}
with the following eigenvectors 
\begin{eqnarray}
  \label{eq:mixc12}
|(1S2p){1/2}^{'-}\rangle&=&-0.334|j=3/2\rangle+0.943|j=1/2\rangle,\cr 
|(1S2p){1/2}^{-}\rangle&=&0.925|j=3/2\rangle+0.380|j=1/2\rangle.
\end{eqnarray}
For the $ccq$ baryon with $J=3/2$ the mixing matrix is given by
\begin{equation}
  \label{eq:mmc32}
  \left(
    \begin{array}{cc}
-23.0&18.1\\ 21.3&18.9
    \end{array}\right)\ {\rm MeV},
\end{equation}
and the eigenvectors are equal to
\begin{eqnarray}
  \label{eq:mixc32}
|(1S2p){3/2}^{'-}\rangle&=&0.343|j=3/2\rangle+0.939|j=1/2\rangle,\cr 
|(1S2p){3/2}^{-}\rangle&=&0.919|j=3/2\rangle-0.394|j=1/2\rangle.
\end{eqnarray}
For the $bbq$ baryon we get the mixing matrix for $J=1/2$
\begin{equation}
  \label{eq:mmb12}
  \left(
    \begin{array}{cc}
-18.0&-2.4 \\-2.8 &-12.6
    \end{array}\right)\ {\rm MeV},
\end{equation}
which has eigenvalues
\begin{eqnarray}
  \label{eq:mixb12}
|(1S2p){1/2}^{'-}\rangle&=&0.349|j=3/2\rangle+0.937|j=1/2\rangle,\cr 
|(1S2p){1/2}^{-}\rangle&=&0.915|j=3/2\rangle+0.402|j=1/2 \rangle,
\end{eqnarray}
and for $J=3/2$ the mixing matrix is
\begin{equation}
  \label{eq:mmb32}
  \left(
    \begin{array}{cc}
-7.4&5.9 \\7.1 &6.3
    \end{array}\right)\ {\rm MeV},
\end{equation} 
so that
\begin{eqnarray}
  \label{eq:mixb32}
|(1S2p){3/2}^{'-}\rangle&=&0.341|j=3/2\rangle+0.940|j=1/2 \rangle,\cr 
|(1S2p){3/2}^{-}\rangle&=&0.917|j=3/2\rangle+0.400|j=1/2 \rangle.
\end{eqnarray}

\begin{figure}
  \begin{turn}{-90} 
\epsfxsize=10cm 
\epsfbox[96 81 518 711]{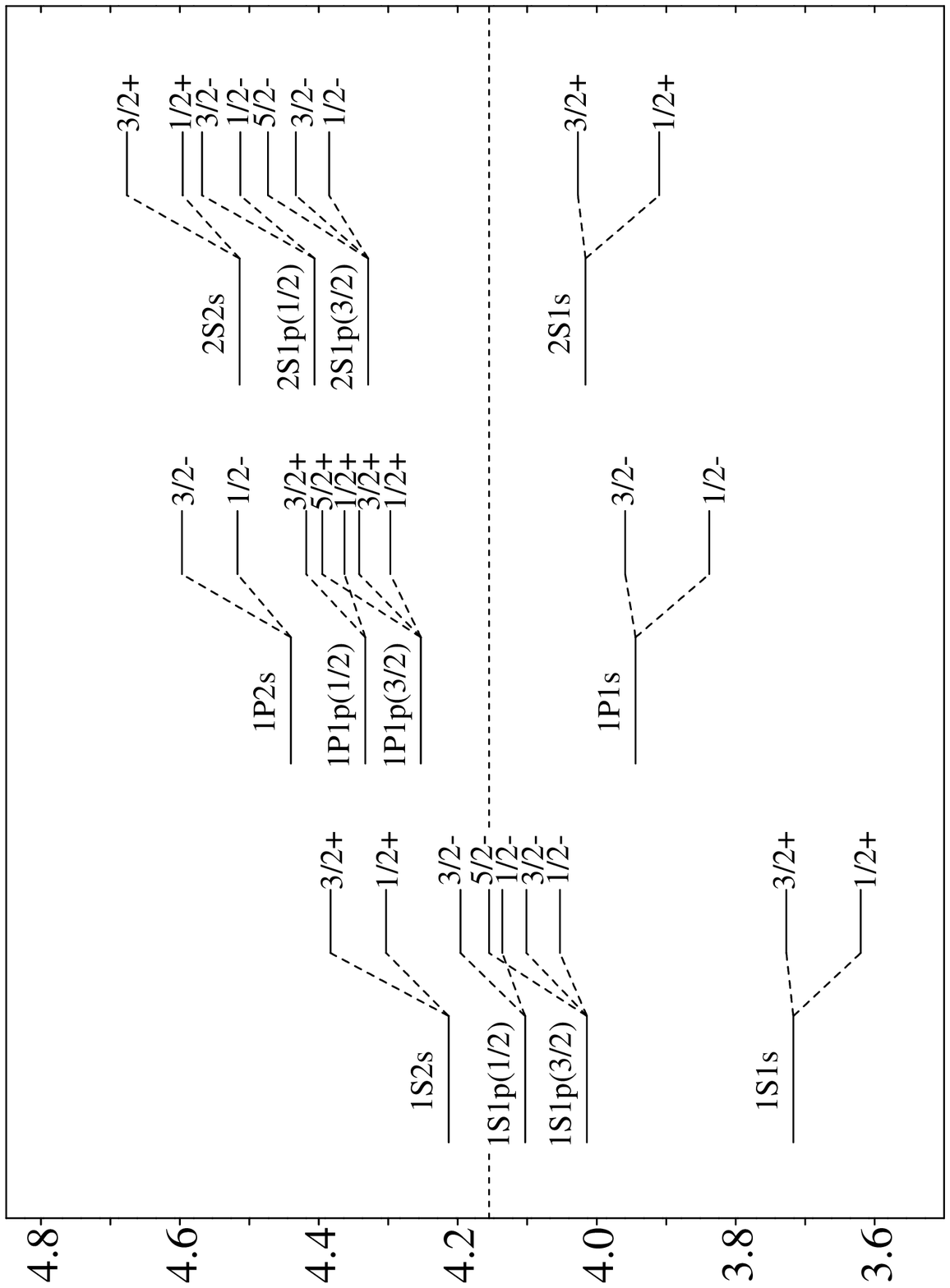}
\end{turn}
\caption{Masses of $\Xi_{cc}$ baryons (in GeV). The horizontal dashed
  line shows the  $\Lambda_c D$ threshold.}
  \label{fig:ccq}
\end{figure}

\begin{figure}
  \begin{turn}{-90} 
\epsfxsize=10cm 
\epsfbox[96 81 518 711]{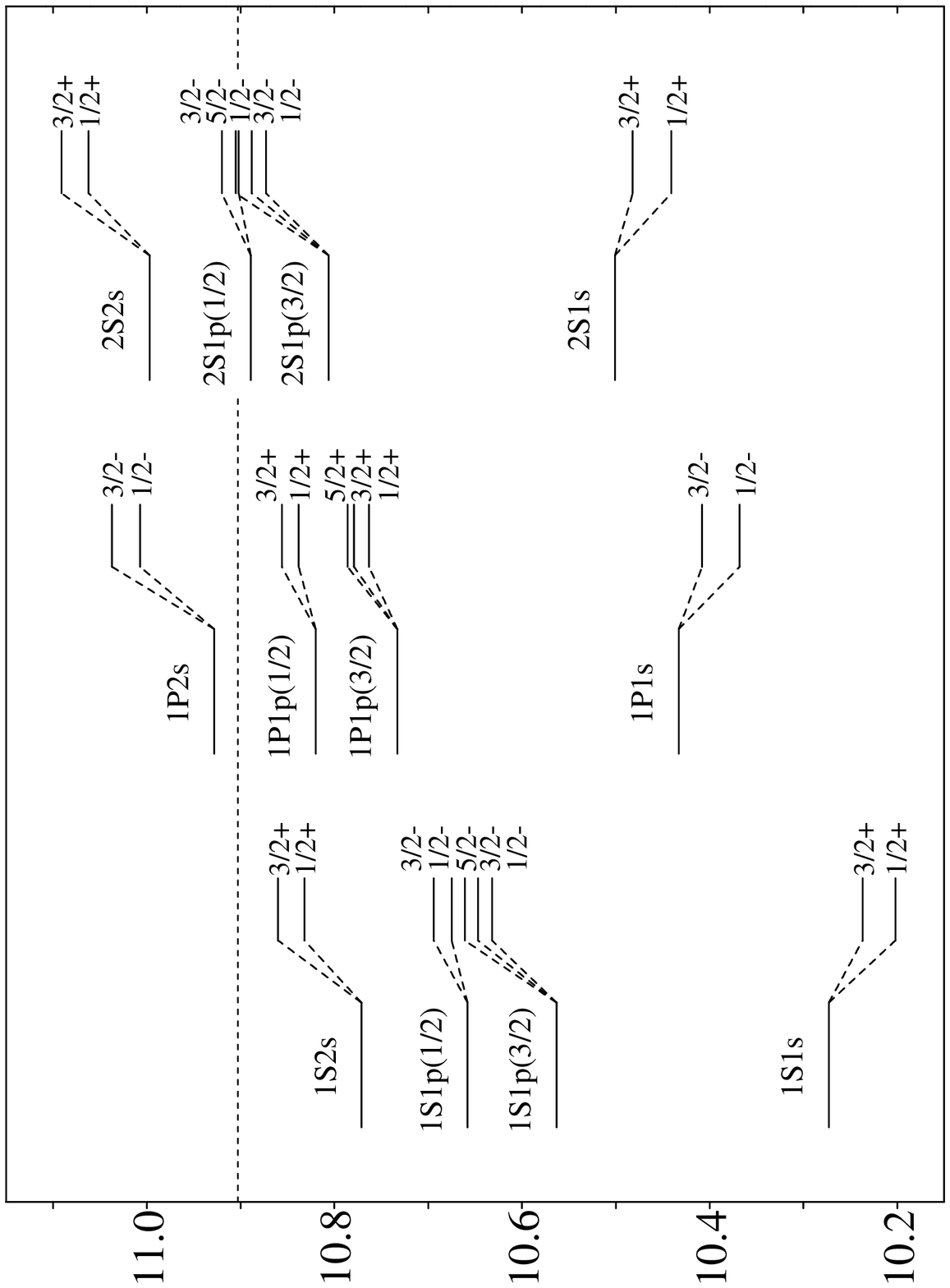}
\end{turn}
\caption{Masses of $\Xi_{bb}$ baryons (in GeV). The horizontal dashed
  line shows the  $\Lambda_b D$ threshold.}
  \label{fig:bbq}
\end{figure}

\begin{figure}
  \begin{turn}{-90} 
\epsfxsize=10cm 
\epsfbox[96 81 518 711]{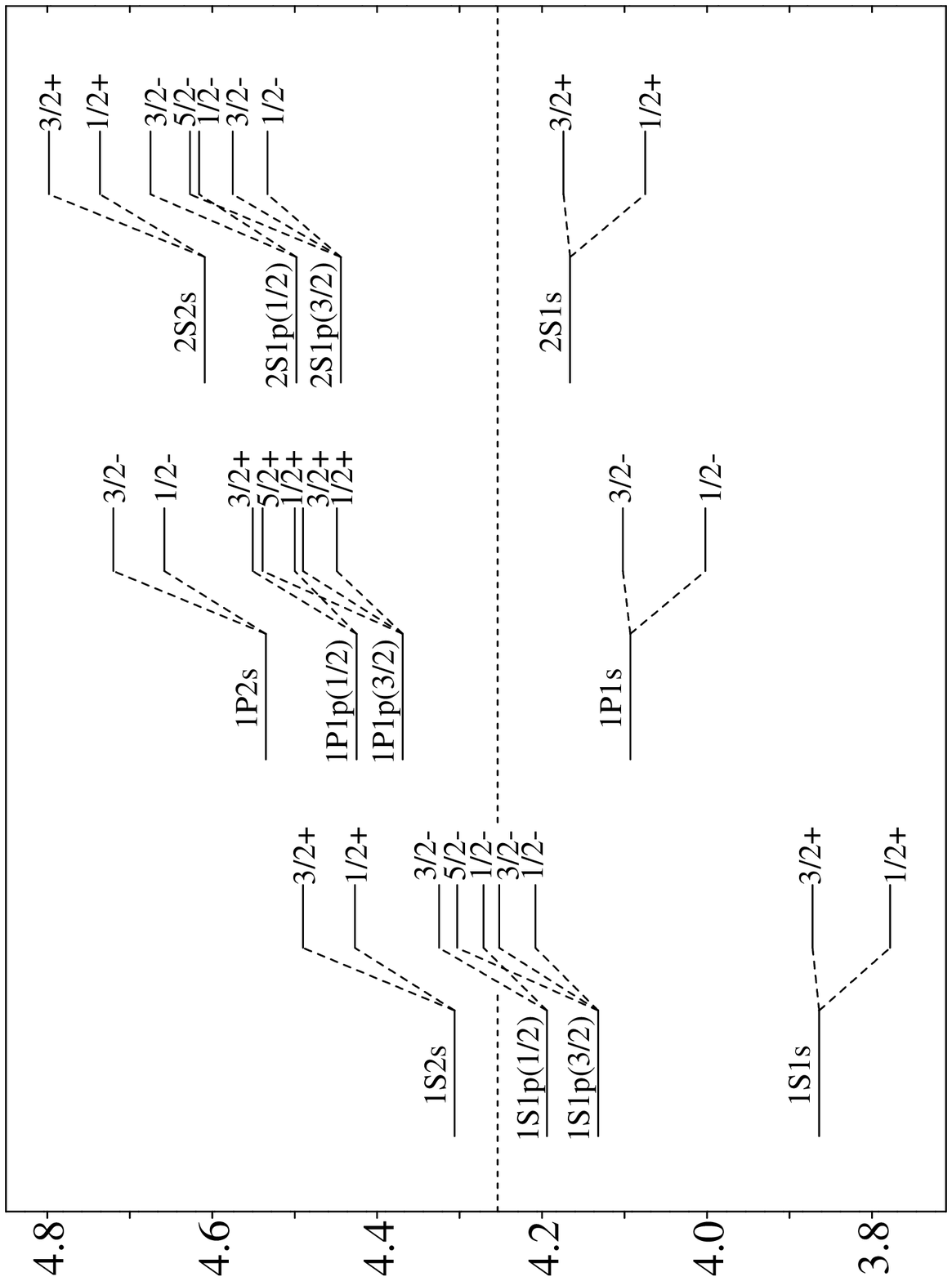}
\end{turn}
\caption{Masses of $\Omega_{cc}$ baryons (in GeV). The horizontal dashed
  line shows the  $\Lambda_c D_s$ threshold.}
  \label{fig:ccs}
\end{figure}

\begin{figure}
  \begin{turn}{-90} 
\epsfxsize=10cm 
\epsfbox[96 81 518 711]{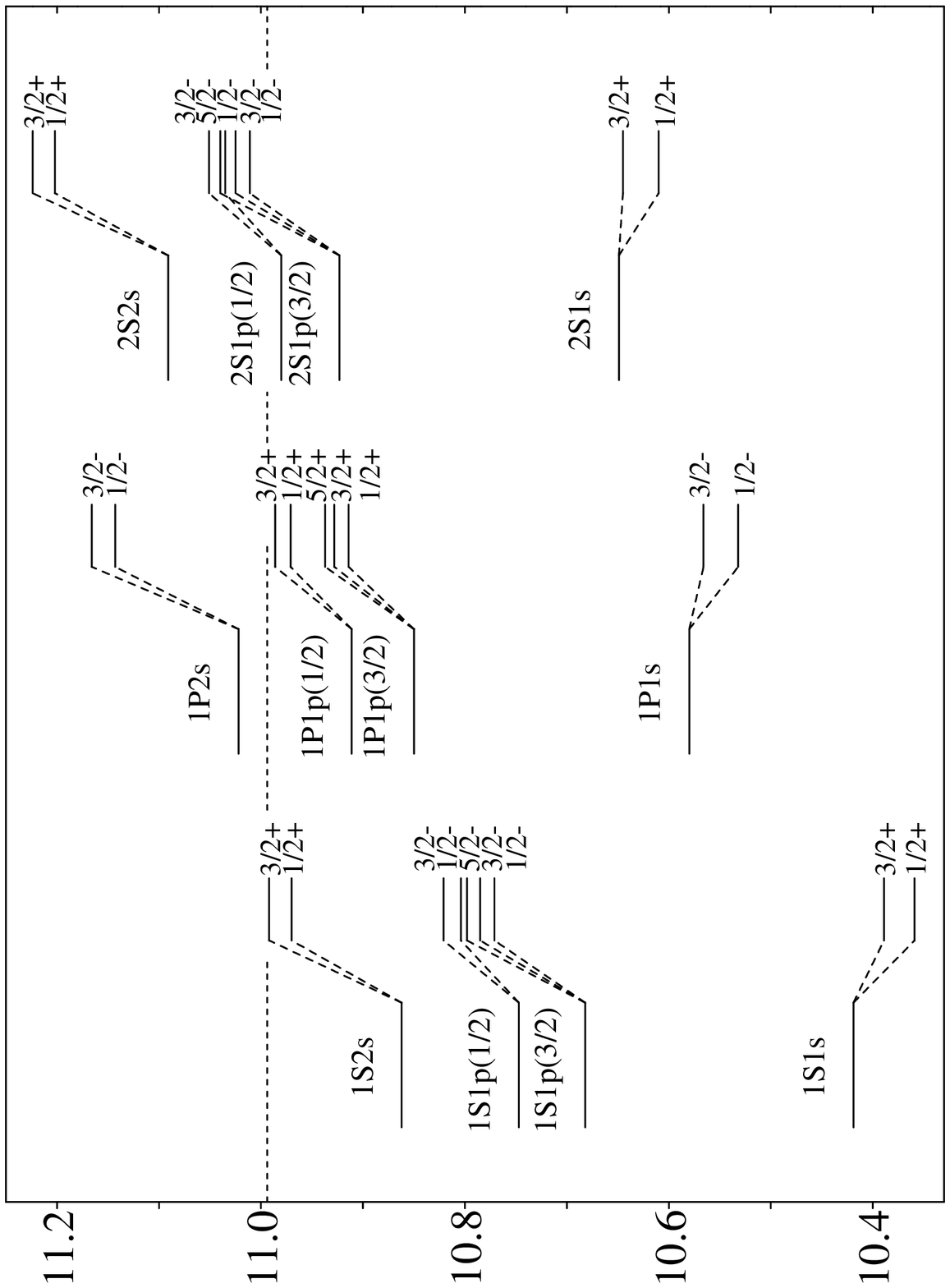}
\end{turn}
\caption{Masses of $\Omega_{bb}$ baryons (in GeV). The horizontal dashed
  line shows the  $\Lambda_b D_s$ threshold.}
  \label{fig:bbs}
\end{figure}

The quasipotential with $1/m_Q$ corrections is
given by the sum of $V_{m_Q\to \infty}(r)$
from (\ref{vinf}) and $\delta V_{1/m_Q}(r)$ from (\ref{svcor}) and
(\ref{avcor}).  By 
substituting it in the quasipotential equation (\ref{quas}) and 
treating the $1/m_Q$ correction term $\delta V_{1/m_Q}(r)$ using
perturbation theory, we are now able to  calculate the
mass spectra of $\Xi_{cc}$, $\Xi_{bb}$,  $\Xi_{cb}$, $\Omega_{cc}$,
$\Omega_{bb}$, $\Omega_{cb}$ baryons with the account of $1/m_Q$ corrections.
In Tables~\ref{tab:ccqmass}--\ref{tab:bbsmass} we present mass spectra
of ground and excited states of doubly heavy baryons containing both
heavy quarks of the same flavour
($c$ and $b$) . The corresponding level orderings are
schematically shown in Figs.~\ref{fig:ccq}--\ref{fig:bbs}. In these
figures we first show our predictions for doubly-heavy baryon spectra
in the limit when all $1/M_d$ corrections are neglected [denoting
baryon states by $n_d L n_q l(j)$]. We see that in this limit the
$p$-wave excitations of the light quark are inverted. This means that
the mass of the state with higher angular momentum $j=3/2$ is
smaller than the mass of the state with lower angular momentum $j=1/2$
\cite{s,egf,isgur}. The similar $p$-level inversion was found
previously in the mass spectra of
heavy-light mesons in the infinitely heavy quark limit \cite{egf}. 
Note that the pattern of levels of the light quark and level
separation in doubly heavy baryons and heavy-light mesons almost
coincide in these limits. 
Next we switch on $1/M_d$ corrections. This results in splitting of
the degenerate states and mixing of states with different $j$, which
have the same total angular momentum $J$ and parity, as it was discussed
above. Since the diquark has spin one, the states with $j=1/2$ split
into two different states with $J=1/2$ or $3/2$, while the states with
$j=3/2$ split into 
three different states with $J=1/2,\ 3/2$ or $5/2$. The fine splitting
between $p$-levels turns out to be of the same order of magnitude as
the gap between $j=1/2$ and $j=3/2$ degenerate multiplets in the infinitely
heavy diquark limit. The inclusion of
$1/M_d$ corrections leads also to the relative shifts of the baryon
levels further decreasing this gap.  
As a result, some of the $p$-levels from different (initially
degenerate) multiplets overlap; however, the heavy diquark spin
averaged centers remain inverted. The resulting picture for the
diquark in the ground state is very similar to the one for heavy-light
mesons \cite{egf}. The purely inverted pattern of $p$-levels  is
observed only for the $B$ meson and $\Xi_{bb}$, $\Omega_{bb}$ baryons,
while in other heavy-light mesons ($D$, $D_s$, $B_s$) and doubly heavy
baryons ($\Xi_{cc}$, $\Omega_{cc}$) $p$-levels from different $j$
multiplets overlap. The absence of the $p$-level overlap  for the
$\Omega_{bb}$ baryon in contrast to the $B_s$ meson (where we predict a
very small overlap of these levels \cite{egf}) is explained by the
fact that the ratio $m_s/M^d_{bb}$ is approximately two times smaller
than $m_s/m_b$ and thus it is of order $m_q/m_b$. As it was argued in
\cite{egf}, these ratios determine the applicability of
the heavy quark limit.

\begin{table}
\caption{\label{tab:ccqmass}Mass spectrum of $\Xi_{cc}$ baryons (in GeV).}
\begin{ruledtabular}
\begin{tabular}{cccccc}
State &  \multicolumn{2}{c}{Mass}  & State & \multicolumn{2}{c}{Mass} \\
$(n_dLn_ql)J^P$& our & \cite{gklo}&$(n_dLn_ql)J^P$& our &
\cite{gklo}\\ 
\hline
$(1S1s)1/2^+$& 3.620& 3.478&$(1P1s)1/2^-$  &3.838   &3.702   \\
$(1S1s)3/2^+$& 3.727& 3.61 &$(1P1s)3/2^-$  &3.959   &3.834    \\
$(1S1p)1/2^-$& 4.053& 3.927&$(2S1s)1/2^+$  &3.910   & 3.812  \\
$(1S1p)3/2^-$& 4.101& 4.039&$(2S1s)3/2^+$  &4.027   &3.944   \\
$(1S1p)1/2'^-$& 4.136& 4.052&$(2P1s)1/2^-$  &4.085   &3.972   \\
$(1S1p)5/2^-$& 4.155& 4.047& $(2P1s)3/2^-$ &4.197   &4.104   \\
$(1S1p)3/2'^-$& 4.196& 4.034&$(3S1s)1/2^+$  &4.154   & 4.072  \\
\end{tabular}
\end{ruledtabular}
\end{table}
\begin{table}
\caption{\label{tab:bbqmass}Mass spectrum of $\Xi_{bb}$ baryons (in GeV).}
\begin{ruledtabular}
\begin{tabular}{cccccc}
State &  \multicolumn{2}{c}{Mass}  & State & \multicolumn{2}{c}{Mass} \\
$(n_dLn_ql)J^P$& our & \cite{gklo}&$(n_dLn_ql)J^P$& our &
\cite{gklo}\\ 
\hline
$(1S1s)1/2^+$& 10.202& 10.093 &$(2S1s)1/2^+$  &10.441   &10.373   \\
$(1S1s)3/2^+$& 10.237& 10.133 &$(2S1s)3/2^+$  &10.482   &10.413    \\
$(1S1p)1/2^-$& 10.632& 10.541 &$(2S1p)1/2^-$  &10.873   &  \\
$(1S1p)3/2^-$& 10.647& 10.567 &$(2S1p)3/2^-$  &10.888   &   \\
$(1S1p)5/2^-$& 10.661& 10.580 &$(2S1p)1/2'^-$ &10.902   &   \\
$(1S1p)1/2'^-$& 10.675& 10.578&$(2S1p)5/2^-$  &10.905   &      \\
$(1S1p)3/2'^-$& 10.694& 10.581&$(2S1p)3/2'^-$ &10.920   &   \\
$(1S2s)1/2^+$& 10.832&        &$(2P1s)1/2^-$  &10.563   &10.493  \\
$(1S2s)3/2^+$& 10.860&        &$(2P1s)3/2^-$  &10.607   &10.533    \\
$(1P1s)1/2^-$& 10.368& 10.310 &$(3S1s)1/2^+$  &10.630   &10.563  \\
$(1P1s)3/2^-$& 10.408& 10.343 &$(3S1s)3/2^+$  &10.673   &   \\
$(1P1p)1/2^+$& 10.763&        &$(3P1s)1/2^-$  &10.744   &  \\
$(1P1p)3/2^+$& 10.779&        &$(3P1s)3/2^-$  &10.788   &    \\
$(1P1p)5/2^+$& 10.786&        &$(4S1s)1/2^+$  &10.812   &   \\
$(1P1p)1/2'^+$& 10.838&       &$(4S1s)3/2^+$  &10.856   &    \\
$(1P1p)3/2'^+$&10.856 &       &$(4P1s)1/2^-$  &10.900   &    \\
\end{tabular}
\end{ruledtabular}
\end{table}
\begin{table}
\caption{\label{tab:ccsmass}Mass spectrum of $\Omega_{cc}$ baryons (in GeV).}
\begin{ruledtabular}
\begin{tabular}{cccc}
$(n_dLn_ql)J^P$& Mass &$(n_dLn_ql)J^P$& Mass\\
\hline
$(1S1s)1/2^+$& 3.778& $(1P1s)1/2^-$  &4.002    \\
$(1S1s)3/2^+$& 3.872& $(1P1s)3/2^-$  &4.102     \\
$(1S1p)1/2^-$& 4.208& $(2S1s)1/2^+$  &4.075      \\
$(1S1p)3/2^-$& 4.252& $(2S1s)3/2^+$  &4.174     \\
$(1S1p)1/2'^-$& 4.271&$(2P1s)1/2^-$  &4.251      \\
$(1S1p)5/2^-$& 4.303& $(2P1s)3/2^-$ &4.345    \\
$(1S1p)3/2'^-$& 4.325&$(3S1s)1/2^+$  &4.321     \\
\end{tabular}
\end{ruledtabular}
\end{table}
\begin{table}
\caption{\label{tab:bbsmass}Mass spectrum of $\Omega_{bb}$ baryons (in GeV).}
\begin{ruledtabular}
\begin{tabular}{cccccc}

$(n_dLn_ql)J^P$& Mass &$(n_dLn_ql)J^P$& Mass\\ 
\hline
$(1S1s)1/2^+$& 10.359&  $(2S1s)1/2^+$  &10.610   \\
$(1S1s)3/2^+$& 10.389&  $(2S1s)3/2^+$  &10.645    \\
$(1S1p)1/2^-$& 10.771&  $(2S1p)1/2^-$  &11.011    \\
$(1S1p)3/2^-$& 10.785&  $(2S1p)3/2^-$  &11.025    \\
$(1S1p)5/2^-$& 10.798&  $(2S1p)1/2'^-$ &11.035      \\
$(1S1p)1/2'^-$& 10.804& $(2S1p)5/2^-$  &11.040        \\
$(1S1p)3/2'^-$& 10.821& $(2S1p)3/2'^-$ &11.051      \\
$(1S2s)1/2^+$& 10.970&  $(2P1s)1/2^-$  &10.738     \\
$(1S2s)3/2^+$& 10.992&  $(2P1s)3/2^-$  &10.775      \\
$(1P1s)1/2^-$& 10.532&  $(3S1s)1/2^+$  &10.806     \\
$(1P1s)3/2^-$& 10.566&  $(3S1s)3/2^+$  &10.843      \\
$(1P1p)1/2^+$& 10.914&  $(3P1s)1/2^-$  &10.924     \\
$(1P1p)3/2^+$& 10.928&  $(3P1s)3/2^-$  &10.961       \\
$(1P1p)5/2^+$& 10.937&  $(4S1s)1/2^+$  &10.994      \\
$(1P1p)1/2'^+$& 10.971& $(4S1s)3/2^+$  &11.031       \\
$(1P1p)3/2'^+$&10.986 & $(4P1s)1/2^-$  &11.083       \\
\end{tabular}
\end{ruledtabular}
\end{table}

In Tables~\ref{tab:ccqmass} and \ref{tab:bbqmass} we compare our
predictions for the ground and excited state masses of $\Xi_{cc}$ and
$\Xi_{bb}$ baryons with the predictions of Ref.~\cite{gklo}. As we see
from these Tables our predictions are approximately $50-150$~MeV
higher than the estimates of Ref.~\cite{gklo}. One of the reasons for the
difference between these two predictions for the masses of $\Xi_{cc}$
baryons (which is the largest) is the difference in the $c$ quark
masses. The mass of the $c$ quark in \cite{gklo} is determined form fitting 
the charmonium spectrum in the quark model where all spin-independent
relativistic corrections were ignored. However, our estimates show
that due to the rather large average value of $v^2/c^2$ in charmonium
\footnote{The spin-dependent relativistic corrections, which are of the
  same order in $v^2/c^2$, produce the level splittings of order of 120
  MeV.} such corrections play an important role and give
contributions to the charmonium masses of order of 100~MeV. As a result
the $c$ quark mass found in \cite{gklo} is approximately 70~MeV less
than in our 
model. For the calculation of the diquark masses we also take into
account the spin-independent corrections (\ref{sipot}) to the $QQ$
potential. We find that their contribution is less than in the case of
charmonium since $V_{QQ}=V_{Q\bar Q}/2$. Thus the $cc$
diquark masses in \cite{gklo} are approximately 50~MeV smaller than in our
model. The other main source of the difference is the expansion in inverse
powers of the light quark mass, which was used in \cite{gklo} but is not
applied in our approach, where the light quark is treated fully
relativistically. 

In Table~\ref{tab:gmass} we compare our model
predictions for the ground state masses of doubly heavy baryons with
some other predictions \cite{gklo,rdlp,kkp,nt} as well as our previous
prediction \cite{efgms}, where the expansion in inverse powers of the heavy
and light quark masses was used. In general we find a reasonable
agreement within 100~MeV between different predictions
\cite{gklo,rdlp,kkp,nt,efgms} for the ground state masses of the doubly
heavy baryons. The main advantage of our present approach is the
completely 
relativistic treatment of the light quark and account for the nonlocal
composite structure of the diquark.                  

For the $\Xi_{cb}$ and $\Omega_{cb}$ baryons containing heavy quarks
of different flavours ($c$ and $b$) we calculate only the ground state
masses. As it was argued in Ref.~\cite{gklo}, the excited
states of heavy diquarks composed of the quarks with different flavours
are unstable under the emission of soft gluons, and thus the
calculation of the excited baryon ($cbq$ and $cbs$) masses is not
justified in the quark-diquark scheme. We get the following predictions
for the masses of the ground state $cbq$ baryons:

$(1S1s)1/2^+$ states with the axial vector and scalar $cb$ diquarks
respectively  
$$M(\Xi_{cb})=6.933~{\rm GeV}, \qquad  M(\Xi'_{cb})=6.963~{\rm GeV},$$

$(1S1s)3/2^+$ state
$$ M(\Xi^*_{cb})=6.980~{\rm GeV},$$
and for $cbs$ baryons:

$(1S1s)1/2^+$ states with the axial vector and scalar $cb$ diquarks
$$M(\Omega_{cb})=7.088~{\rm GeV}, \qquad  M(\Omega'_{cb})=7.116~{\rm GeV},$$

$(1S1s)3/2^+$ state
$$ M(\Omega^*_{cb})=7.130~{\rm GeV}.$$

\begin{table}
\caption{\label{tab:gmass}Mass spectrum of ground states of doubly
  heavy baryons (in GeV). Comparison of different
  predictions. $\{QQ\}$ denotes the diquark in the axial vector state
  and $[QQ]$ denotes diquark in the scalar state.}
\begin{ruledtabular}
\begin{tabular}{ccccccccc}
Baryon&
Quark&$J^P$&Present&\cite{gklo}&\cite{efgms}&\cite{rdlp}&\cite{kkp}&
\cite{nt}\\
      &content&    &work&     &&&\\   
\hline
$\Xi_{cc}$  &$\{cc\}q$&$1/2^+$&3.620&3.478&3.66&3.66&3.61&3.69\\
$\Xi_{cc}^*$&$\{cc\}q$&$3/2^+$&3.727&3.61&3.81&3.74&3.68&\\
$\Omega_{cc}$&$\{cc\}s$&$1/2^+$&3.778&3.59&3.76&3.74&3.71&3.86\\
$\Omega_{cc}^*$&$\{cc\}s$&$3/2^+$&3.872&3.69&3.89&3.82&3.76&\\
$\Xi_{bb}$  &$\{bb\}q$&$1/2^+$&10.202&10.093&10.23&10.34& &10.16\\
$\Xi_{bb}^*$ &$\{bb\}q$&$3/2^+$&10.237&10.133&10.28&10.37& &\\
$\Omega_{bb}$&$\{bb\}s$&$1/2^+$&10.359&10.18&10.32&10.37& &10.34\\
$\Omega_{bb}^*$&$\{bb\}s$&$3/2^+$&10.389&10.20&10.36&10.40& &\\
$\Xi_{cb}$  &$\{cb\}q$&$1/2^+$&6.933&6.82&6.95&7.04& &6.96\\
$\Xi'_{cb}$  &$[cb]q$&$1/2^+$&6.963&6.85&7.00&6.99& &\\
$\Xi_{cb}^*$ &$\{cb\}q$&$3/2^+$&6.980&6.90&7.02&7.06& &\\
$\Omega_{cb}$  &$\{cb\}s$&$1/2^+$&7.088&6.91&7.05&7.09& &7.13\\
$\Omega'_{cb}$  &$[cb]s$&$1/2^+$&7.116&6.93&7.09&7.06& &\\
$\Omega_{cb}^*$ &$\{cb\}s$&$3/2^+$&7.130&6.99&7.11&7.12& &\\
\end{tabular}
\end{ruledtabular}
\end{table}

Now we compare our results with the model-independent predictions of
the heavy quark effective theory.
The heavy quark symmetry predicts simple relations between the spin
averaged masses of doubly heavy baryons with the accuracy of order
$1/M_d$
\begin{eqnarray}
  \label{eq:sad}
  \Delta \bar M_{1,2}&\equiv&\bar M_1(\Xi_{bb})-\bar M_1(\Xi_{cc})= \bar
  M_2(\Xi_{bb})-\bar M_2(\Xi_{cc})= \bar M_1(\Omega_{bb})-\bar
  M_1(\Omega_{cc})\cr
  &=& \bar M_2(\Omega_{bb})-\bar M_2(\Omega_{cc})=
  M^d_{bb}-M^d_{cc}\equiv \Delta M^d,
\end{eqnarray}
where the spin-averaged masses are 
\begin{eqnarray*}
  \bar M_1&=&(M_{1/2}+2M_{3/2})/3,\cr
  \bar M_2&=&(M_{1/2}+2M_{3/2}+3M_{5/2})/6
\end{eqnarray*}
and $M_J$ are the masses of baryons with total angular momentum $J$.
$M^d_{QQ}$ are the masses of diquarks in definite states. The
numerical results are presented in Table~\ref{tab:samd} (only the
states below threshold are considered). We see that the equalities in
Eq.~(\ref{eq:sad}) are satisfied with good accuracy  for the baryons
with the heavy diquark and light quark both in ground and excited states.  
\begin{table}
\caption{\label{tab:samd}Differences between spin-averaged masses of
  doubly heavy baryons defined in Eq.~(\ref{eq:sad}) (in GeV).}
\begin{ruledtabular}
\begin{tabular}{cccccc}
Baryon state& $\Delta\bar M_1(\Xi)$&$\Delta\bar M_2(\Xi)$
&$\Delta\bar M_1(\Omega)$&$\Delta\bar M_2(\Omega)$& $\Delta M^d$\\
\hline
$1S1s$& 6.534&      & 6.538&         &6.552\\
$1S1p$& 6.512& 6.532&      &6.519    &6.552\\
$1P1s$& 6.476&      & 6.486&         &6.484\\
$2S1s$& 6.480&      & 6.492&         &6.480
\end{tabular}
\end{ruledtabular}
\end{table} 

It follows from the heavy quark symmetry that
the hyperfine mass splittings of initially degenerate light quark
states 
\begin{eqnarray}
 \Delta M(\Xi_{QQ})&\equiv& M_{3/2}(\Xi_{QQ})-M_{1/2} (\Xi_{QQ}),\cr
\Delta M(\Omega_{QQ})&\equiv& M_{3/2}(\Omega_{QQ})-M_{1/2}
(\Omega_{QQ}),
\end{eqnarray}
should scale with the diquark masses:
\begin{eqnarray}
  \Delta M(\Xi_{cc})&=& R \  \Delta M(\Xi_{bb}),\cr
\Delta M(\Omega_{cc})&=& R\ \Delta M(\Omega_{bb}),
\end{eqnarray}
where $R=M_{bb}^d/M_{cc}^d$ is the ratio of diquark masses. Our model
predictions for these splittings are displayed in
Table~\ref{tab:hfs}. Again we see that heavy quark symmetry relations
are satisfied with high accuracy.
\begin{table}
\caption{\label{tab:hfs} Hyperfine splittings (in MeV) of the doubly
  heavy baryons for the states with the light quark angular momentum
  $j=1/2$.}
\begin{ruledtabular}
\begin{tabular}{cccccccc}
Baryon state&$R$&$\Delta M(\Xi_{bb})$ &$R\ \Delta
M(\Xi_{bb})$& $\Delta M(\Xi_{cc})$ &$\Delta M(\Omega_{bb})$&
$R\ \Delta M(\Omega_{bb})$&$\Delta M(\Omega_{cc})$ \\
\hline
$1S1s[3/2-1/2]$&3.03  & 35 & 106& 107  & 30  &91& 94\\
$1S1p[3/2-1/2]$& 3.03 & 19 & 58 & 60  & 17  &52 & 54\\
$1P1s[3/2-1/2]$& 2.87 & 40 & 115& 121 & 34  &98& 100\\
$2S1s[3/2-1/2]$& 2.83 & 41 & 116& 117  & 35  &99& 99
\end{tabular}
\end{ruledtabular}
\end{table} 

The close similarity of the interaction of the light quark with the
heavy quark in the heavy-light mesons and with the heavy diquark in
the doubly heavy baryons produces very simple relations between the
meson and baryon mass splittings \cite{sw,lmw,l}. In fact for the ground
state hyperfine splittings of mesons and baryons we obtain adopting the
approximate relation $M_{QQ}^d\cong 2 m_Q$
\begin{eqnarray}
  \label{eq:mbr}
  \Delta M(\Xi_{QQ})&=&\frac32\frac{m_Q}{M_{QQ}^d}\Delta M_{B,D} \cong
  \frac34 \Delta M_{B,D},\cr
  \Delta M(\Omega_{QQ})&\cong&\frac34 \Delta M_{B_s,D_s},
\end{eqnarray}
where the factor $3/2$ is just the ratio of the baryon  and
meson spin matrix elements. The numerical fulfillment of
relations (\ref{eq:mbr}) is shown in Table~\ref{tab:hfsmb}. 
\begin{table}
\caption{\label{tab:hfsmb}Comparison of hyperfine splittings (in MeV) in doubly
  heavy baryons and heavy-light mesons. Experimental values
  for the hyperfine splittings in mesons are taken from
  Ref.~\cite{pdg}.} 
\begin{ruledtabular}
\begin{tabular}{cccccccc}
$\Delta M(\Xi_{cc})$& $\frac34 \Delta M_D^{\rm exp}$& $\Delta M(\Xi_{bb})$&
$\frac34 \Delta M_B^{\rm exp}$ & $\Delta M(\Omega_{cc})$& $\frac34 \Delta
M_{D_s}^{\rm exp}$ & $\Delta M(\Omega_{bb})$& $\frac34 
\Delta M_{B_s}^{\rm exp}$\\
\hline
107& 106& 35 &34 &94 &108 & 30 &35 
\end{tabular}
\end{ruledtabular}
\end{table}

\section{Conclusions} 
\label{sec:concl}

In this paper we calculated the masses of the ground and 
excited states of the doubly heavy baryons on the basis of the
quark-diquark approximation in the framework of the relativistic quark model.
The orbital and radial excitations both of the heavy diquark and the light
quark were considered. The main advantage of the proposed approach consists
in the fully relativistic treatment of the light quark ($u,d,s$)
dynamics and in the account for the internal structure of the diquark
in the short-range quark-diquark interaction. We apply only the
expansion in inverse powers of the heavy diquark mass
($M_{bb}^d,M_{cc}^d$), which considerably simplifies calculations. The
infinitely heavy 
diquark limit as well as the first order $1/M_d$ spin-independent and
spin-dependent contributions were considered. A close similarity between
excitations of the light quark in the doubly heavy baryons and
heavy-light mesons was demonstrated. In the infinitely heavy (di)quark limit
the only difference originates from the internal structure of the diquark
which is important at small distances. The first order contributions
to the heavy (di)quark expansion explicitly depend on the 
values of the heavy diquark (boson) and heavy quark (fermion) spins and
masses ($M^d_{QQ}\approx 2m_Q$). This results in the
different number of levels to which the initially degenerate states
split as well as their ordering.  
Our model respects the constraints imposed by heavy quark symmetry on
the number of levels and their splittings.       

We find that the $p$-wave levels of the light quark which correspond to
heavy diquark spin multiplets with $j=1/2$ and $j=3/2$  are inverted in
the infinitely heavy diquark limit.
The origin of this inversion is the following. The confining potential
contribution to the spin-orbit term in (\ref{vinf}) exceeds the one-gluon
exchange contribution. Thus the sign before the spin-orbit term is negative,
and the level inversion emerges. However, the $1/M_d$ corrections, which
produce the hyperfine splittings of these multiplets, are
substantial. As a result the purely inverted pattern of $p$ levels
for the heavy diquark in the ground state occurs only for the doubly
heavy baryons $\Xi_{bb}$ and $\Omega_{bb}$.
For $\Xi_{cc}$ and $\Omega_{cc}$ baryons the levels from these
multiplets overlap. The similar pattern was previously found
in our model for the heavy-light mesons \cite{egf}.

We plan to use the found wave functions of doubly heavy baryons  for the
calculation of semileptonic and nonleptonic $\Xi_{bb}$ decays to
$\Xi_{cb}$ and $\Xi_{cb}$ to $\Xi_{cc}$ baryons. The corresponding
baryonic Isgur-Wise functions \cite{kkp,efkr} will be determined.

\acknowledgments
The authors express their gratitude to M. M\"uller-Preussker,
V. Savrin and H. Toki  for support and discussions.
Two of us (R.N.F and V.O.G.) were supported in part by the 
{\it Deutsche Forschungsgemeinschaft} under contract Eb 139/2-1 and
{\it Russian Foundation for Basic Research} under Grant
No.~00-02-17768. The work of R.N.F, V.O.G. and A.P.M. was supported in 
part by  {\it Russian Ministry of Education} under Grant No.~E00-3.3-45.


\end{document}